\documentclass{www13-companion-accepted}

\usepackage{color}
\usepackage{url}

\usepackage{booktabs}
\usepackage{ctable}
\usepackage{multirow} 
\usepackage{rotating}
\usepackage{tabulary}

\usepackage[justification=centering]{caption}
\usepackage{subfigure}
\usepackage{array}
\usepackage{varwidth}
\usepackage[normalem]{ulem}
\usepackage{enumitem}

\usepackage{balance}

\usepackage[bottom, hang, flushmargin, ragged]{footmisc}

\usepackage{xfrac}
\usepackage{amsmath}

\newcolumntype{P}[1]{>{\centering\raggedright\arraybackslash}p{#1}}

\newcommand{\otoprule}{\midrule[\heavyrulewidth]}

\begin{document}

\conferenceinfo{WWW}{'13 Rio de Janeiro, Brazil}

\title{RESLVE: Leveraging User Interest to Improve \\Entity Disambiguation on
Short Text}
%
%
%
%
%

\numberofauthors{3} 
%
\author{
%
%
\alignauthor Elizabeth L. Murnane\\
\affaddr{Cornell University\\Information Science}\\
\email{elm236@cornell.edu}
\alignauthor Bernhard Haslhofer\\
\affaddr{Cornell University\\Information Science}\\
\email{bh392@cornell.edu}
\alignauthor Carl Lagoze\\
\affaddr{University of Michigan\\School of Information}\\
\email{clagoze@umich.edu}
}

\maketitle

\begin{abstract}

We address the Named Entity Disambiguation (NED) problem for short,
user-generated texts on the social Web. In such settings, the lack of linguistic
features and sparse lexical context result in a high degree of ambiguity and
sharp performance drops of nearly 50\% in the accuracy of conventional NED
systems. We handle these challenges by developing a model of user-interest with
respect to a personal knowledge context; and Wikipedia, a particularly
well-established and reliable knowledge base, is used to instantiate the
procedure. We conduct systematic evaluations using individuals' posts from
Twitter, YouTube, and Flickr and demonstrate that our novel technique is able to
achieve substantial performance gains beyond state-of-the-art NED methods.

\end{abstract}

\category{H.3.3}{Information Storage and Retrieval}{Information Search and
Retrieval}
\category{I.2.7}{Artificial Intelligence}{Natural Language
Processing}

\terms{Algorithms, Design, Experimentation, Human Factors}

\keywords{Entity Resolution; Social Web; Semantic Knowledge Graph;
User Interest Modeling; Personalized IR}

\section{Introduction}
Named Entity Recognition (NER) refers to the systematic process of identifying
mentions of \textit{entities} such as people, places, concepts, or ideas in
unstructured text. The Named Entity Disambiguation (NED) problem arises when an
entity has multiple \textit{candidate} meanings, and it is particularly
challenging when texts are short, linguistic features are unreliable, and local
lexical context is sparse. This kind of scenario is epitomized by the highly
individualized text-based utterances found in online social media platforms,
where the amount of such content is continually expanding and the need to
extract information and knowledge from it grows.

The following examples illustrate the difficulty in determining the unique,
intended sense of an ambiguous entity.\\[5pt] A tweet, posted on Twitter, the
fastest growing micro-blog\-ging service, where the amount of tweets produced
in a day is equivalent to a 10 million page book\footnote{\url{http://blog.twitter.com/2011/06/}}:\\[2pt]
\centerline{\textit{aaahh one more day until \uline{\textbf{finn!!!}}
\#cantwait}}\\[4pt] A video title posted on YouTube, a video sharing website
with over 800 million visitors each month\footnote{\url{http://www.youtube.com/t/press_statistics}}:\\[2pt]
\centerline{\textit{the \uline{\textbf{office}} holiday party}}\\[5pt] A photo
tag on Flickr, which hosts over 7 billion images, a figure nearly double the
amount from just 4 years ago\footnote{\url{http://blog.flickr.net/en}}:\\[2pt]
\centerline{\textit{ \uline{\textbf{Beetle}}}}\\[5pt] According to their
corresponding Wikipedia Disambiguation pages, these entities have dozens of
candidate meanings. For instance, ``finn" is both a popular TV character or a
travel destination; ``office'' might refer to the concept of a workplace,
an American television show, or a British television show; and ``Beetle'' could
mean the animal, one of several vehicle models, a botanist, or a simple dice
game.

In each of these cases, the lexical context in which the ambiguous entity is
contained does not help to definitively determine the user's intended meaning
or is absent completely. Attempting to utilize prior tweets, videos, or photos
posted by the user in the same social Web platform as a source of background knowledge is not an effective
approach either, both because users do not generally post a large enough
volume and because the ambiguity in those posts is often equally
as high, offering no informative context \cite{Michelson:2010jo}.

Thus even though NED is a well-established problem with much extant research
devoted to it, this short text written on social media is
fundamentally different than the long and formal text on which traditional
approaches are trained \cite{Inches:2010uz}; and consequently, unique
identification of entity concepts is severely degraded for even state-of-the-art
conventional methods \cite{Meij:2012bd}.
For instance, the F1 score of the Stanford NER, which is trained on the
CoNLL-03\footnote{\url{http://www.cnts.ua.ac.be/conll2003}} news article
dataset, drops from over 90\% to just under 46\% when applied to a Twitter
dataset \cite{Liu:2011uf}.

With researchers beginning to recognize this, the area of entity
recognition within social media has begun to draw some recent attention. Several
strategies use crowd-sourcing techniques for linking entities, but these methods
require a large number of reliable human workers
\cite{Demartini:2012bt, Finin:2010wd}. Automated approaches to NER
have emerged, but they also face limitations. Namely, most only handle entity extraction without addressing the disambiguation problem, and
nearly all have only been trained to handle tweets, making it unclear whether
results are generalizable to personal utterances outside of Twitter
\cite{Davis:2012vx, Li:2012gm, Liu:2011uf, Ritter:2011vb}.

Our work moves beyond these strategies and establishes 
individual-centric procedures to automatically disambiguate the short yet
personally relevant text-based utterances a user makes on a variety of platforms. Our
solution compensates for sparse lexical context with \textit{personal context},
building a model of user interest from external structured semantic data.
Essentially, we propose that an ambiguous entity's intended meaning is the
candidate concept most similar to a user's core set of personal interests.

We introduce a novel system called RESLVE (\textbf{R}esolving \textbf{E}ntity
\textbf{S}ense by \textbf{L}e\textbf{V}eraging \textbf{E}dits) that augments
traditional disambiguation techniques by implementing this personalized
approach. RESLVE relies on state-of-the-art services such as Wikipedia Miner
\cite{Milne:2008wu} and DBPedia Spotlight \cite{Mendes:2011bi} for extraction of entities and candidate meanings, and our improvements to the
disambiguation phase can increase the overall effectiveness of such tools. The
central contributions of this paper are:
\begin{itemize}[itemsep=3pt,topsep=3pt,parsep=1pt,partopsep=0pt,leftmargin=2em]
  \item A model for representing user interest with respect to a knowledge base
  and its categorical organization scheme.
  \item A ranking technique that takes as input a user interest model along with
  an ambiguous entity's set of candidate topics and outputs with improved
  accuracy the topic most likely intended.
  \item An annotated dataset of disambiguated entities from Twitter, YouTube,
  and Flickr along with the results of an empirical evaluation of our
  system.
\end{itemize}
The dataset and system implementation are publicly available at
\url{https://github.com/emurnane/RESLVE}.

The remainder of the paper is organized as follows. Section~\ref{sec:modeling_user_interest} presents the
theoretical underpinnings of our approach to model user interest according to
the text-based traces people leave on the Web. Section~\ref{sec:reslve_system}
details our implementation of the model as well as our ranking technique that
measures the relevance of an ambiguous entity's candidate topics with respect to this model.
Section~\ref{sec:experiments} describes the experiment we performed to evaluate
the method, offering insights and statistics about our datasets and details of data
preparation. Section~\ref{sec:results} presents results, comparing our
performance to baseline methods and NED services and identifying cases of
efficacy and error. Finally, Section~\ref{sec:conclusion} offers concluding
remarks and directions for future research.

\section{Modeling User Interest}
\label{sec:modeling_user_interest}
\subsection{Intuitions and Illustrations}
The underlying assumptions of our approach are that a user possesses a core set
of interests, a user is more likely to mention an entity about a topic drawn
from a domain of personal interest than from a domain of non-interest, and these
interests can be formally modeled as a personal knowledge context. Thus by
bridging user identity between social media (e.g., Twitter, YouTube, or Flickr)
and a knowledge base (e.g., Wikipedia), we propose to resolve the intended
meaning of ambiguous entities encountered in the former by leveraging the power
of structured information relevant to personal interests available in the
latter.

To aid understanding of these ideas, we present qualitative
examples based on the Web traces of users in our dataset. Earlier, we gave an
example of an ambiguous YouTube title: ``\textit{the \textbf{\uline{office}}
holiday party}''. Without seeing the video, even
a human is unable to determine from title alone whether the video contains
footage of a workplace, TV episode, or some other
office-related subject matter; and extra detail like ``episode 4'' still does
not aid manual disambiguation between the US and UK shows.
Inspecting the user's YouTube profile, we find other video
titles mentioning this entity, such as ``\textit{the \textbf{\uline{office}}, december 3}'', but that information does not help in figuring out the meaning of the entity in question either.
Examining the user's log of recent Wikipedia edits, however, allows us to resolve at
once that the likely meaning is the US TV show:
\textit{<item userid=xx user=xx pageid=31841130 title=\textbf{\uline{The Office (U.S. season 8)}}/>}

Next, a key assumption is that the interests expressed by a user on various Web
platforms do not wander considerably among an incoherently diverse set of topics
but rather are relatively consistent. To quantify this extent to which user
interests overlap across platforms, we evaluate the coverage that Wikipedia
categories\footnote{\url{http://en.wikipedia.org/wiki/Special:Categories}}
provide for topics mentioned in users' social Web utterances. We find that on
average within our dataset, 54.2\% of the entities a user mentions in social
media (e.g., ``Java'') have one or more of their candidate meanings covered by a
direct category of an article that same user edited on Wikipedia (e.g.,
``Programming\_language'').
\begin{figure}[b]
	\centering
	\setlength\fboxsep{0pt}
	\setlength{\fboxrule}{1pt}
	\framebox{%
    	\includegraphics[width=0.6\linewidth]{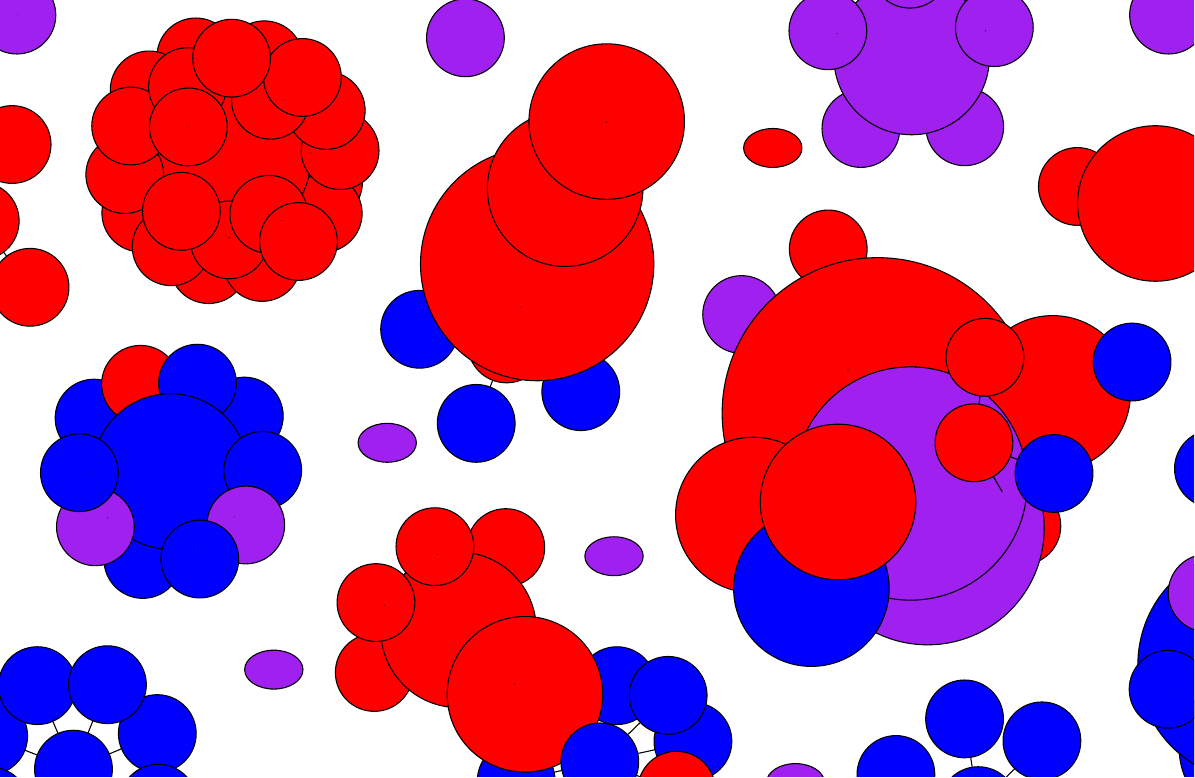}%
	}
	\caption{Example of the overlap in topics a user utters on Twitter (blue), on
	Wikipedia (purple), and on both platforms (red)}
	\label{fig:interest_coverage}
\end{figure}
Figure~\ref{fig:interest_coverage} provides an illustrative example of a
representative user's overlap of topics mentioned on both Twitter and Wikipedia.
Each circle represents a topic with size proportional to the number of times
the user uttered or edited that topic; and topics are grouped according to their
proximity in Wikipedia's category hierarchy. For instance, a majority of this
user's tweets and edited articles deal with information technology concepts,
frequently overlapping on topics of programming languages, internet browsers,
and security vulnerabilities. In fact, of the 178 entities this user mentions in
tweets, 74.7\% of them have a candidate meaning belonging to a direct category
of an article she contributed to in Wikipedia (i.e., a category 1 edge away in
the category hierarchy), 90.4\% belong to a category 2
edges away, 98.9\% to a category 3 edges away, and all 100\% of the entities
mentioned in tweets are covered by moving only 4 edges up the category hierarchy
starting at some article the user edited.

We now buttress these intuitions by looking both to theory and
validated experimental results from prior research.
\subsection{Theoretical Motivations}
Firstly, our work builds on social psychology theories that say a
user possesses intrinsic interest in a key set of topics. Expressing these
interests is personally fulfilling, motivating consumption and contribution of
content about those topics to online communities \cite{Lakhani:2003ep,
Lerner:2000tc, Maslow:1970vl}, especially compared to random topics; for
instance, people are more likely to read and respond to posts mentioning movies
of personal interest than to random posts \cite{Harper:2007:TAY:1216295.1216313,
Ling:2006bx}.
In addition, prior research shows that a user's online contributions in social
media and knowledge production communities are representative of that
individual's main topics of interest, and this research further establishes that
these interests can be modeled according to lexical features of her text-based
contributions \cite{Chen:2010er, Cosley:2007kr, Pennacchiotti:2011wh}. The next
section describes how we represent user interest with respect to a general knowledge base.

\subsection{Modeling a Knowledge Context}
\label{sec:general_model}
Put simply, a \textit{knowledge base} is a semantic network used to organize
entities, their types, and the relations among them. Modeling topics with respect to this structured information
allows us to formally represent the topics in which a user is interested and the
topics corresponding to an ambiguous entity's candidate meanings. In Section~\ref{sec:reslve_system}, we
instantiate our devised model using Wikipedia.

\begin{figure}[b]
	\centering
	\includegraphics[width=0.7\linewidth]{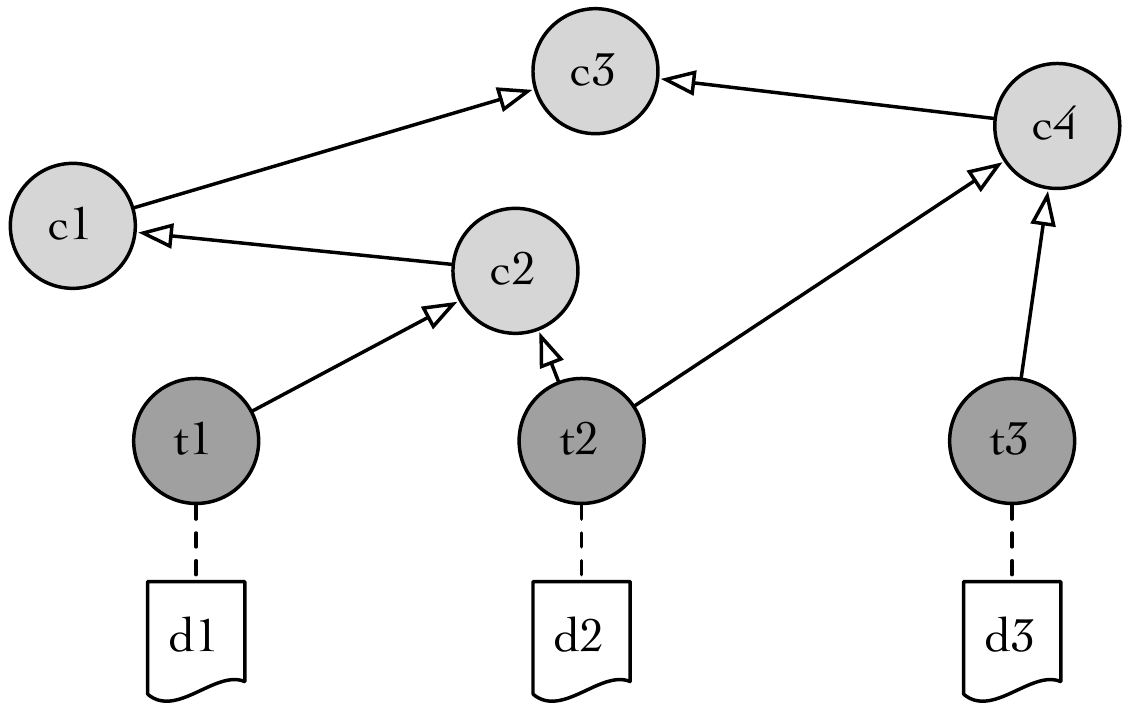}
	\caption{Knowledge graph of categories, topics, and descriptions}
	\label{fig:knowledge_graph}
\end{figure}

We begin with some definitions. First, we define a knowledge base $K$ as a
directed graph $K=(N,E)$, consisting of sets of nodes $N$ and edges $E$, as
illustrated in Figure~\ref{fig:knowledge_graph}. We distinguish between two types of
nodes:

\textit{Category nodes}. $N_{Category} \subset{N}$ with each
$c\in{N_{Category}}$ having a unique identifier $i\in{I}$ and a set of semantic
relationships $r$ with other nodes, so that $c = \{i,R\}$. As an example, a
category Car has a $rel:broader$ relationship to category Vehicle, and both
category Vehicle and category Elephant have $rel:broader$ relationships to
category Thing.

\textit{Topic nodes}. $N_{Topic}\subset{N}$ with each $t\in{N_{Topic}}$ also
having a unique identifier $i\in{I}$, belonging to one or more categories, and
carrying a textual description $d$ so that $t = \{i,d,C\}$ with
$C\subseteq{N_{Category}}$. For example, Volkswagen Beetle is a topic belonging
to category Car and having an associated descriptive document, which in the case
of Wikipedia could be the text on the Volkswagen Beetle article page.

With this definition of a knowledge graph in hand, we can build the interest
model of a user $u$. To begin, we treat the user's text-based contributions to
the description of a topic $t$ as a signal of interest in that topic. For each
such topic $t_{i}\in{T_{u}}$, where $T_{u} \subseteq{N_{Topic}}$ is the set of
all topics a user has contributed to the description of, we construct a
topic-interest graph rooted at node $t_{i}$ and consisting of all categories
$C_{u} \subseteq{N_{Category}}$ that are reachable from $t_{i}$ by following
outgoing edges to parent categories in the knowledge graph $K$.
(Note that we represent a candidate meaning of an ambiguous entity in the same
way; the only difference is that a candidate is associated with the single topic
that captures its meaning, while the representation of a user contains each
topic to which that user has contributed). Formally, we represent user interest
$I$ in topic $t_{i}$ as:\\
\centerline{$I_{t_{i}}=(N_{u},E_{u})$, where $N_{u_{i}}=t_{i} \cup C_{u}$ and
$E_{u_{i}} \subseteq E$}

We ensure each $U_{t_{i}}$ is a bipartite graph by applying the following
straightforward transformation. We initialize all edge weights in
$U_{t_{i}}$ to 1. Then, starting at node ${t_{i}}$, we traverse each outgoing
edge $e_{(i,j)}=(n_{i},n_{j})$ until
we encounter an edge between two categories, that is an edge where both
$n_{i},n_{j}\in{C_{u}}$, at which point we apply the transformation:\\[2pt]
a. Compute $p$ = shortest path length between $t_{i}$ and $n_{j}$\\[2pt]
b. Remove the edge $e_{(i,j)}$ and replace it with an edge from $t_{i}$ to
$n_{j}$ that has an edge weight equal to $\frac{1}{p}$.  If such an edge
$e_{(t_{i},n_{j})}$ already exists, assign it a new weight equal to the
greater of its current weight and $\frac{1}{p}$.
Thus the interest model of user $u$ is simply an aggregated set of all
topic-interest graphs built from each topic the user has shown direct or
indirect interest in, formally defined as
$u=\{I_{t_{1}},I_{t_{2}},\ldots,I_{t_{n}}\}$.

\begin{figure}[b]
	\centering
	\includegraphics[width=0.75\linewidth]{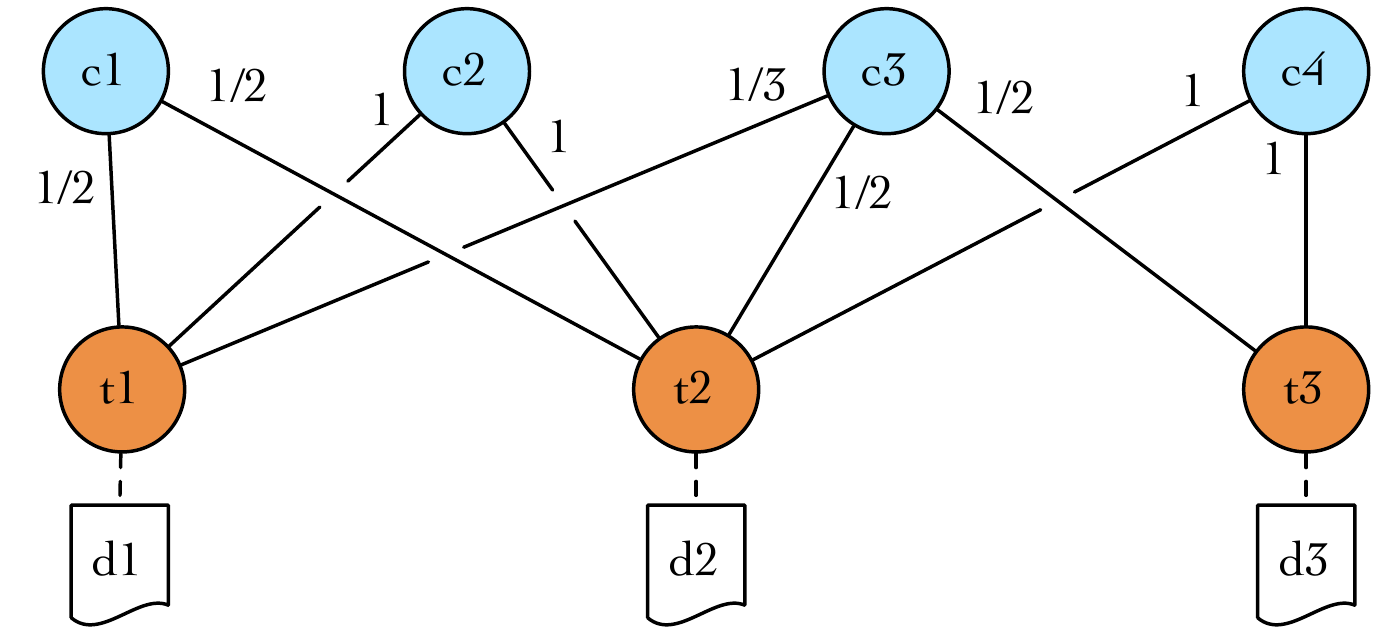}
	\caption{User topic-interest graph}
	\label{fig:interest_graph}
\end{figure}

To summarize, we now have a
consolidated user-interest graph consisting of all topic nodes a user has
contributed to, connected to category nodes by edges weighted with a value
$w, 0<w\leq{1}$. As this aggregation of multiple topic-interest graphs may
result in a graph with duplicate categories, we can eliminate duplicates by simply
first removing any category node $c'$ having the same identifier as another
category $c$ in the graph and then taking any edge originally between a topic
and $c'$ and redrawing it to $c$ instead.
To illustrate this process, let a user's set of contributions correspond to
$\{t_{1},t_{2},t_{3}\}$ as shown in Figure~\ref{fig:knowledge_graph}.
Applying the steps to transform the graph into a bipartite graph and removing duplicate category nodes results
in Figure~\ref{fig:interest_graph}.

It is also possible at this point to represent the user interest graph as a
topic-category matrix, which is the analogue of a term-document matrix that
represents documents in a vector space model. Figure~\ref{fig:interest_matrix} illustrates such a matrix built from the
topic-interest graph shown in Figure~\ref{fig:interest_graph}. Each row corresponds to a vector for topic $t_{i}$, each column corresponds to a category
$c\in{C_{u}}$, and each topic-category entry equals the weight of the edge
between those two nodes in the user's interest graph or 0 if no such edge
exists.
\begin{figure}[h]
	\vspace*{-.4cm}
	\[ \left( \begin{array}{cccc}
	\sfrac{1}{2} & 1 & \sfrac{1}{3} & 0 \\
	\sfrac{1}{2} & 1 & \sfrac{1}{2} & 1 \\
	0 & 0 & \sfrac{1}{2} & 1 \end{array} \right)\] 
	\vspace*{-.3cm}
	\centering
	\caption{Edge-weight matrix of user interest graph}
	\label{fig:interest_matrix}
\end{figure}

Having provided a walkthrough of our theoretical motivations and a method for
constructing a model of user interest, we next describe the full details of our
implemented system. We could instantiate our model on any knowledge base with
the above structure such as Wikipedia, DBPedia, and Freebase, which are
all high coverage knowledge bases employed as sources of entity and concept
representations for Twitter \cite{Laniado:2010ih}, YouTube \cite{Batista:2009iz}, and Flickr \cite{Ruocco:2012by}.
For this paper, we implement on Wikipedia for reasons described in
Section~\ref{sec:wikipedia}.
\section{The RESLVE System}
\label{sec:reslve_system}
To determine the intended meaning of an
ambiguous entity detected in a user's short text, we compute the
relevance of each candidate topic to which an entity could refer to the user's
interest model, which is based on the topics that user contributed to in a
knowledge base, for example Wikipedia.
\subsection{Wikipedia as a Knowledge Context}
\label{sec:wikipedia}
Wikipedia is both popular and powerful, capable of facilitating entity
recognition, linking, and disambiguation \cite{Bunescu:2006vx, Cucerzan:2007tb,
Meij:2012bd, Mihalcea:2007hf, Milne:2008wu}. For this paper, we implement our
model using Wikipedia both because it has been well established as a rich source
of external information and because it offers additional advantages particularly relevant for our task.

First, Wikipedia has been shown effective in modeling user interests since
editing behavior on the site serves as an indicator of interest and because the
site's organizational structure provides a way to formally represent those
interests. Specifically, Wikipedia editors seek out
articles about topics of personal interest, revision
histories catalog these topics along with valuable metadata, and resources like
article pages and category graphs effectively represent the topics
\cite{Cosley:2007kr, Lieberman:2009us, Strube:2006wa, Syed:2008wr,
Wattenberg:2007vd}.
\begin{table}[b]
	\caption{Editing behaviors indicative of user interest}
	\vspace*{-.4cm}
	\begin{tabular}{p{2.9cm} p{4.6cm}}
		\toprule
		\multicolumn{1}{c}{\textbf{Editing Behavior}}
		 & \multicolumn{1}{c}{\textbf{Intuition}}
		\tabularnewline \toprule
		Number of times user edits article & 
		Repeated editing implies greater investment and interest
		\\ \midrule
		Type of edit & Trivial edits (see Table~\ref{tab:ignored_edits}) are weaker
		signal of interest
		\\ \midrule
		Article's global edit activity and number of editors & Generally
		popular articles are less discriminative of individual interest and personal relevance
		\\ \midrule
		Editing time span & Long-term interests are stronger than fleeting ones \\
		\midrule
		Edit quality w.r.t. Info. Qual. metrics & Substantiveness and quality indicate
		concern in topic \\ \bottomrule	
	\end{tabular}
	\label{tab:editing_behavior}
\end{table}
In addition, research analyzing social media shows that users frequently mention
topics of personal interest in tweets and that Wikipedia's category structure
can be used to represent those topics \cite{Michelson:2010jo}. Also, Wikipedia
provides broad coverage of domain-independent named entity concepts and rare
word senses \cite{Li:2012gm, Zesch:2007vs}, which is key given the diversity of
topics users talk about on the social Web.

Finally, article-editing behavior lends itself well to formulating the strength
of a user's interest in a given topic.
Table~\ref{tab:editing_behavior} summarizes contribution characteristics that
are indicative of interest and can be incorporated into a weighting formula,
along with underlying intuitions as to why such behavior is meaningful. For this
experiment, we factor in only the number of edits and the edit type.

Some work that Wikipedia users perform, however, is not indicative of personal
interest \cite{Cosley:2007kr} and is therefore filtered out when constructing
the interest model. The left column of Table~\ref{tab:ignored_edits} summarizes
such irrelevant edits, and the table's right column lists the further processing
we perform to clean all article text.
\begin{table}[t]
	\caption{Edits and text that are not concept-bearing}
	\vspace*{-.4cm}
	\begin{tabular}{p{3cm} p{4.5cm}}
		\toprule
		\multicolumn{1}{c}{\textbf{Edits Ignored:}} &
		\multicolumn{1}{c}{\textbf{Patterns Cleaned:}}\\
		$\bullet$ Trivial (typo fixes, &  $\bullet$ Stopwords, punct. removed \\
		\ \ \ vandalism reverts) & $\bullet$ Article maintenance info in \\
		$\bullet$ Articles with under & \ \ \ Wiki Markup removed \\
		\ \ \ 100 non-stopwords  &$\bullet$ Stem, tokenize, lowercase \\ \bottomrule	
	\end{tabular}
	\label{tab:ignored_edits}
\end{table}

To reiterate, our approach simply takes each Wikipedia article a user has made
non-trivial contributions to and constructs an interest model according to the
specification described in Section~\ref{sec:general_model}. That is, we treat
articles as the \textbf{\textit{topics}} comprising $N_{Topic}$, use Wikipedia
``Category:" resources to represent the
\textbf{\textit{categories}} in $N_{Category}$, and consider the text content on an article page to be that topic's unique \textbf{\textit{description}}.

\begin{figure*}
	\centering
	\includegraphics[scale=0.7]{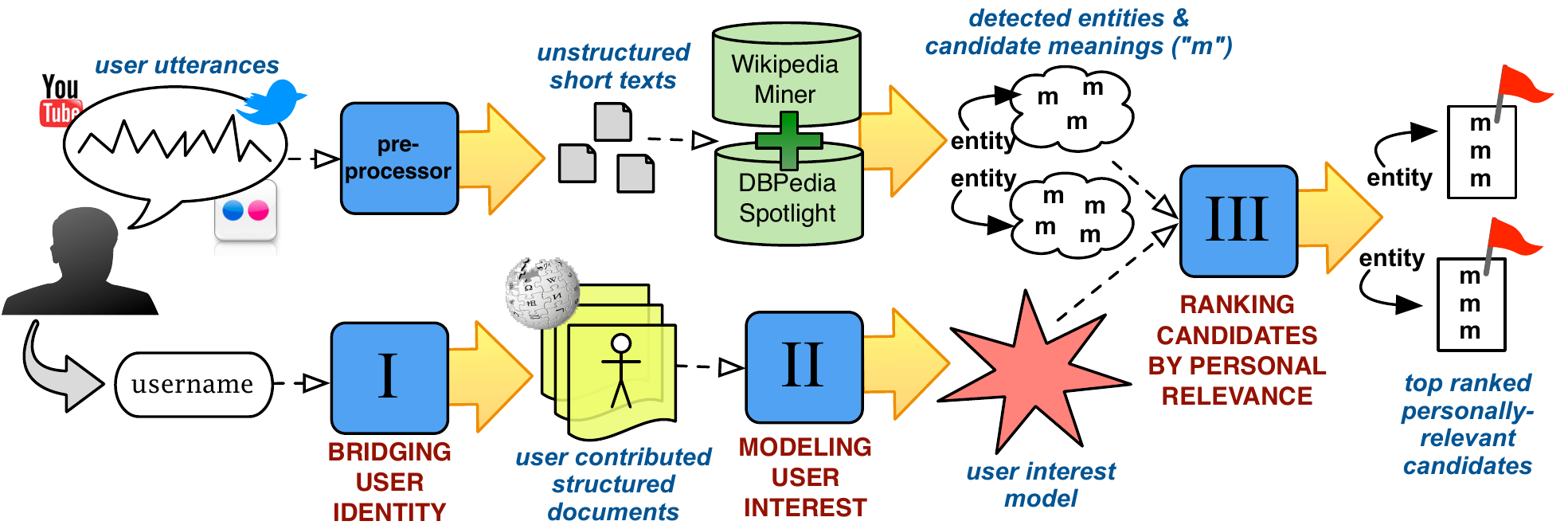}
	\caption{The RESLVE system performs disambiguation by (I) connecting a
	user's social web identity and Wikipedia editor identity, (II) modeling
	personal interests in terms of topics associated with the articles the user
	edited on Wikipedia, and (III) ranking entity candidates by measuring how similar each candidate's topic is to
	the user's interest model}
	\label{fig:system_framework}
\end{figure*}

We now move on to describe the modules in our system framework, RESLVE, that
addresses the disambiguation problem by (I) connecting a user's social Web identity and Wikipedia editor identity, (II) modeling that
user's personal interests using her articles edited on Wikipedia, and (III)
ranking entity candidates by measuring how similar each candidate's associated
topic is to the most salient topics in the user's interest model.
Throughout our discussion, we refer to Figure~\ref{fig:system_framework},
which illustrates each of those phases.
\subsection{Phase I: Bridging Web Identities}
\label{sec:bridging_approach}
The first step is bridging the user's social Web identity with that same user's
identity in the context of a knowledge base. Given an
ambiguous text on social media, we detect the Wikipedia account
that belongs to the same person to leverage its record of
contributions as structured personal context that can be used to formally model
interests. Our current approach is simple string matching of account usernames
since prior research demonstrates feasibility \cite{Iofciu:2011th,
Perito:1325528}.
See Section~\ref{sec:experiments} for validation of the technique.
\subsection{Phase II: Representing Users and Entities}
\label{sec:vector_rep}
RESLVE next ranks an ambiguous entity's candidate topics by measuring how
relevant each one is to the user's interest model built from the bridged
account.
RESLVE takes advantage of both a topic's description (Section~\ref{sec:content_sim}) as well as its semantic relationships in the
knowledge-graph (Section~\ref{sec:category_sim}) in order to get an overall
measure of relatedness (Section~\ref{sec:combo_sim}) between the user's
interests and each candidate meaning. We represent the user \textbf{$u$} and the
candidate meaning \textbf{$m$} as weighted vectors in order to use classic
information-retrieval techniques to measure their similarity.
\subsubsection{Content-Based Similarity}
\label{sec:content_sim}
To measure content-based similarity, we use a bag-of-words approach, building
both for $u$ and for $m$ a TF-IDF weighted term vector from the titles of the
articles a user has edited, the candidate's article title, the
tokenized words from all those articles' pages, and the titles of those
articles' categories. $sim_{content}$ is the cosine-similarity measurement
between the user vector $V_{content,u}$ and the candidate meaning vector
$V_{content,m}$. Given a total of $\gamma$ terms extracted from articles'
titles, descriptions, and categories, we have:
\begin{align*}
V_{content,u}&=\{tfidf(u,t_{1}),\ldots,tfidf(u,t_{\gamma})\} \\
V_{content,m}&=\{tfidf(m,t_{1}),\ldots,tfidf(m,t_{\gamma})\} \\
sim_{content}(u,m)&=cossim(V_{content,u},V_{content,m})
\end{align*}
\subsubsection{Knowledge-Context Based Similarity}
\label{sec:category_sim}
To measure relevance based on semantic relationships from the knowledge graph,
we build vectors now using articles' category IDs.
$V_{category}=\{w_{graph}(category_{i}) \mid{category_{i}} \in{(C_{u} 
\cup{C_{m})}}, C_{u},C_{m} \subseteq{N}\}$, where each $category_{i}$ is a
category with an edge to an interest topic or a candidate topic. To consider
not only occurrence but position of a category in the knowledge graph, the
function $w_{graph}$ measures a category's ``semantic relevance'' to a topic,
denoted $dist(c)$, as the edge weight between the topic and category 
(the shortest path length as explained in Section~\ref{sec:general_model}). This
scheme assigns more weight to a close and directly relevant category of a topic such as ``American Television Series" than to a category far away and too general such as ``Broadcasting". We
also do factor in the occurrence of a category, denoted $freq(c)$, which is the
number of user- or candidate-relevant articles the category has an edge with in
the knowledge graph. For a candidate, this value is either 0 or 1 depending on
whether the category is present or absent in the hierarchy of categories originating from the candidate's corresponding article.
Formally the weighting formula is defined as:\\
\centerline{$w_{graph}(c_{i},u)  =\left\{
\begin{array}{c l}      
    dist(c_{i}) \ast freq(c_{i}),
    & c_{i} \in{C_{u}}\\
    0 & c_{i} \not\in{C_{u}}
\end{array}\right.$\\}\\
\centerline{$w_{graph}(c_{i},m)=\left\{
\begin{array}{c l}      
    dist(c_{i}) \ast freq(c_{i}),
    & c_{i} \in{C_{m}}\\
    0 & c_{i} \not\in{C_{m}}
\end{array}\right.$\\}\\
\centerline{where $C_{u},C_{m} \subseteq{N}$}\\
The semantic relevance between $u$ and $m$ is then: \\
\centerline{$sim_{category}(u,m)=cossim(V_{category,u},V_{category,m})$}
\subsection{Phase III: Ranking by Personal Relevance}
\label{sec:combo_sim}
All candidate meanings for an ambiguous entity are scored using the
composite formula $sim(u,m)= \alpha \ast{sim_{content}(u,m)}\\ + (1-\alpha)
\ast{sim_{category}(u,m)}$ where $\alpha$ is a weighting parameter determined
experimentally. RESLVE outputs the highest scoring candidate as the user's
intended meaning.

Given that Wikipedia has millions of articles and hundreds of thousands of
categories, it is necessary to avoid vector dimensionalities that make
computation impractical. For the future we will explore a pruning strategy that
requires new vector components to meet a ``relatedness'' threshold to the
current vector in addition to using standard strategies that remove very high
and low frequency terms.
\section{Experiments}
\label{sec:experiments}
\subsection{Data Collection and Preparation}
\label{sec:data_proc}
We collected tweets posted on Twitter, videos posted on YouTube, and photos
posted on Flickr. After eliminating users whose content was not written in
English and the accounts that had since been deleted or had public permissions
removed, the remaining usernames were fed to the identity module (Module I in
Figure~\ref{fig:system_framework}), which output 
usernames that existed on Wikipedia too.
We then employed annotators to confirm each bridged account belonged to one
individual person in order to both assess the effectiveness of our string-matching approach (see Section~\ref{sec:bridging_approach}) and to ensure that the accounts used in the experiment were not false positives that would pollute
results. Section~\ref{sec:bridging_results} summarizes the annotator judgments
in Table~\ref{tab:username_reuse} and offers solutions to overcome any low
username-reuse encountered.

Next, guided by thresholds used in recent research \cite{Hauff:2012es,
Lu:2012tg, Zhang:2012uf}, accounts with less than 100 social Web
utterances or 100 lifetime Wikipedia edits were removed as inactive. We downloaded the
remaining users' most recent 100 short text utterances on each social Web
platform - specifically, their tweets, YouTube video titles and descriptions,
and Flickr photo tags, titles, and descriptions. We also collected the ID, title, page content, and categories for
every article the user edited on Wikipedia in order to build a comprehensive personal context. We
did not consider interest drift here, but we have future plans to investigate
whether temporal attributes of a user's utterances and edits affect the
observable interest overlap.

We passed the short texts to the pre-processing module of
Figure~\ref{fig:system_framework} to be cleaned and filtered in the ways listed
in Table~\ref{tab:text_proc}.
\begin{table}[t]
	\caption{Short text pre-processing and normalization}
	\vspace*{-.4cm}
	\begin{tabular}{@{}p{.8cm}p{6.7cm}}
		\toprule
		\textbf{Tweets:} & 
		\vspace*{-.2cm}
		\begin{itemize}[itemsep=0pt,topsep=0pt,parsep=0.5pt,partopsep=0pt,leftmargin=3em]
  			\item Normalize @name to MENTION
  			\item Remove RT (retweet) tag
  			\item Remove leading ``\#'' but keep hash tag's target concept if English
  			word \end{itemize} \\[-\normalbaselineskip]
		\midrule
		\textbf{YouTube, Flickr:} &
		\vspace*{-.2cm}
		\begin{itemize}[itemsep=0pt,topsep=0pt,parsep=0.5pt,partopsep=0pt,leftmargin=3em]
  			\item Bypass auto-generated file names like IMG\_336.jpg or MOV\_02.AVI
  			\item Remove file type suffix, e.g. ``.png'', but leave file name if an English
  			word
  			\item Ignore auto-generated tags, e.g., ``hidden:filter=Boost''
  			machine-tag on Flickr
		\end{itemize} \\[-\normalbaselineskip]
		\midrule
		\vspace*{-.2cm}
		\textbf{\shortstack[l]{All utter-\\ances:}} &
		\vspace*{-.2cm}
		\begin{itemize}[itemsep=0pt,topsep=0pt,parsep=0.5pt,partopsep=0pt,leftmargin=3em]
  			\item Remove URLs
  			\item Remove non-English
		\end{itemize} \\[-\normalbaselineskip]
		\bottomrule
	\end{tabular}
	\label{tab:text_proc}
\end{table}
RESLVE then extracted entities from the short texts using Wikipedia Miner and
DBPedia Spotlight with disambiguation threshold parameters set to 0 in order to
retrieve all potential candidate meanings. The filters in Table~\ref{tab:entity_filters}
removed invalid entities, and we gave the remaining 1545 valid entities to
Mechanical Turkers tasked with labeling whether each candidate provided by the
NER services was the correct meaning or not. Three different annotators labeled
each candidate, and we required workers to be Categorization
Masters\footnote{\url{https://www.mturk.com/mturk/help?helpPage=worker}} or have
at least 95\% approved hits in prior tasks. We removed entities for which
Turkers did not unanimously select a correct sense, leaving 918 labeled ambiguous entities.
\begin{table}[b]
	\caption{Filters applied to entities}
	\vspace*{-.4cm}
	\begin{tabular}{@{}p{0.75cm}p{6.75cm}}
		\toprule
		\textbf{Language based:} & 
		\vspace*{-.2cm}
		\begin{itemize}[itemsep=0pt,topsep=0pt,parsep=0.5pt,partopsep=0pt,leftmargin=3em]
  			\item Non-English
  			\item Single characters and parse errors
		\end{itemize} \\[-\normalbaselineskip]
		\midrule
		\textbf{Entity based:} &
		\vspace*{-.2cm}
		\begin{itemize}[itemsep=0pt,topsep=0pt,parsep=0.5pt,partopsep=0pt,leftmargin=3em]
  			\item Non-entities, i.e., detected terms that are not a Noun class (NN,
  			NNS, NNP, NP) or Named Entity class (e.g., location, person, organization) according to named entity corpora IEER, ACE, or CoNLL
  			\item Non-ambiguous entities (0 or 1 meaning)
		\end{itemize}  \\[-\normalbaselineskip]
		\bottomrule
	\end{tabular}
	\label{tab:entity_filters}
\end{table}
In the final step, ambiguous entities were passed to module III (see
Figure~\ref{fig:system_framework}), which ranks candidates according to
their similarity with the user interest model and outputs the most likely
intended meaning. We give descriptive statistics about the texts and entities
in Section~\ref{sec:data_characteristics}, and we present RESLVE's ranking 
precision in Section~\ref{sec:results}.
\subsection{Dataset Characteristics}
\label{sec:data_characteristics}
\begin{figure}
	\centering
	\setlength\fboxsep{0pt}
	\setlength{\fboxrule}{1pt}
	\framebox{%
    	\includegraphics[width=0.7\linewidth]{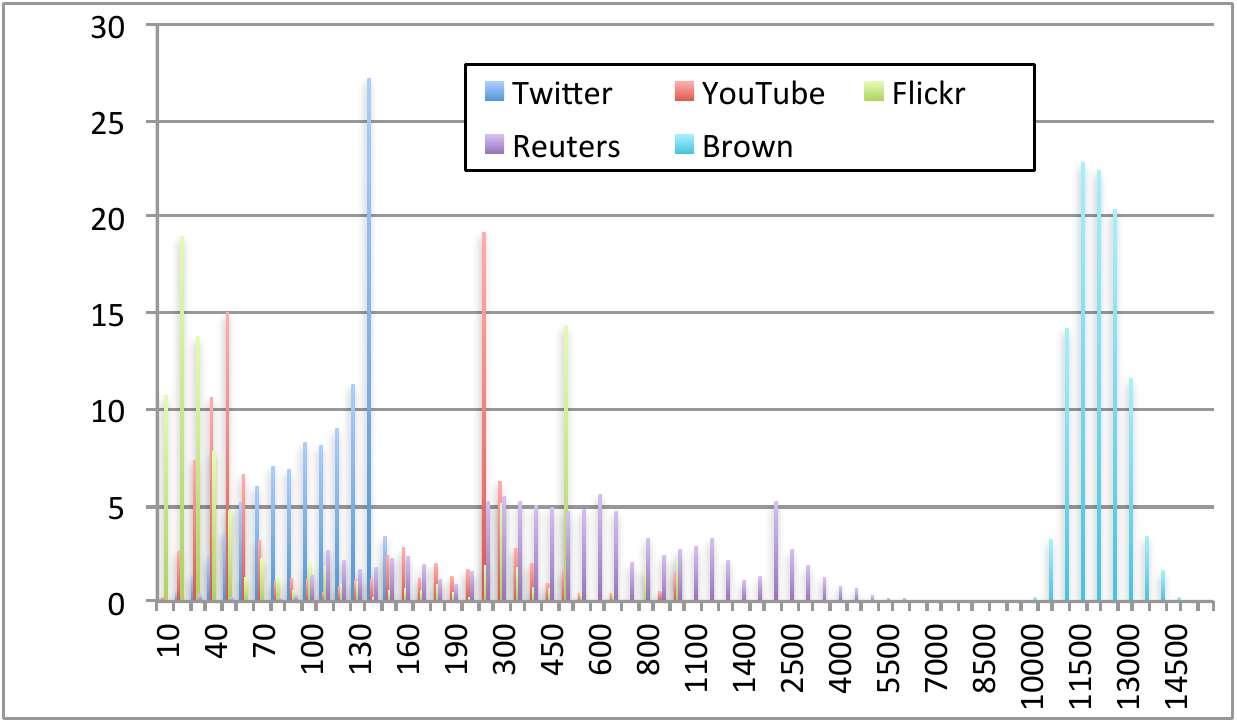}%
	}
	\caption{Freq (y) of chars (x) in various corpora}
	\label{fig:corpora}
\end{figure}

Here we report on interesting attributes and general trends found in our
dataset.
Figure~\ref{fig:corpora} shows the distributions of text length (number of
characters) for utterances on Twitter, YouTube, and Flickr as
well as for content in
Reuters-21578\footnote{\url{http://kdd.ics.uci.edu/databases/reuters21578/reuters21578.html}}
and Brown-Corpus\footnote{\url{http://www.hit.uib.no/icame/brown/bcm.html}}
collections. The last two are corpora on which standard NED tools are
trained and tested. Apparent is the very short nature of social
Web posts, especially photo tags, photo titles, and video titles; and
the longest texts a user writes on the social Web are still generally shorter
than even the shortest texts from traditional NER task corpora.
The character limits imposed on these social platforms also stand out as peaks
in the graph. Tweets can be at most 140 characters; Flickr has a 500-character
limit; and YouTube allows 1000 characters for video descriptions but only
250 show in a preview area, so users' self-limiting causes a noticeable peak
there.

Furthermore, we see the occurrence of entities with multiple meanings is
high. Table~\ref{tab:ambiguity} summarizes the
proportions of tweets, titles, descriptions, and tags that contain at least one ambiguous
entity as well as the percentage of detected entities that are ambiguous. Across
all 3 social Web sites, 91\% of users' content contain one or multiple
ambiguous entities; nearly two-thirds of all detected entities on these sites
are ambiguous; and very few entity-containing short texts contain only non-ambiguous entities.
\begin{table}[h]
	\caption{Ambiguity on social Web. (a): texts containing ambiguous entities;
	(b): entities with ambiguous sense}
	\vspace*{-.35cm}
	\begin{tabular}{c|c|cc|ccc}
		\toprule
		 \multicolumn{1}{c}{} & \multicolumn{1}{c}{\textbf{Twitter}} &
		 \multicolumn{2}{c}{\textbf{YouTube}} & \multicolumn{3}{c}{\textbf{Flickr}}
		 \\ \toprule
		 & \textit{Tweet} & \textit{Title} & \textit{Desc} & \textit{Title} &
		 \textit{Desc} & \textit{Tag} \\
		 (a) & 93\% & 88\% & 98\% & 92\% & 97\% & 77\% \\
		 (b) & 64\% & 55\% & 46\% & 66\% & 44\% & 73\% \\
		 \bottomrule
	\end{tabular}
	\label{tab:ambiguity}
\end{table}
\begin{figure}[b]
	\centering \setlength\fboxsep{0pt}
	\setlength{\fboxrule}{1pt}
\framebox{%
   \includegraphics[width=0.7\linewidth]{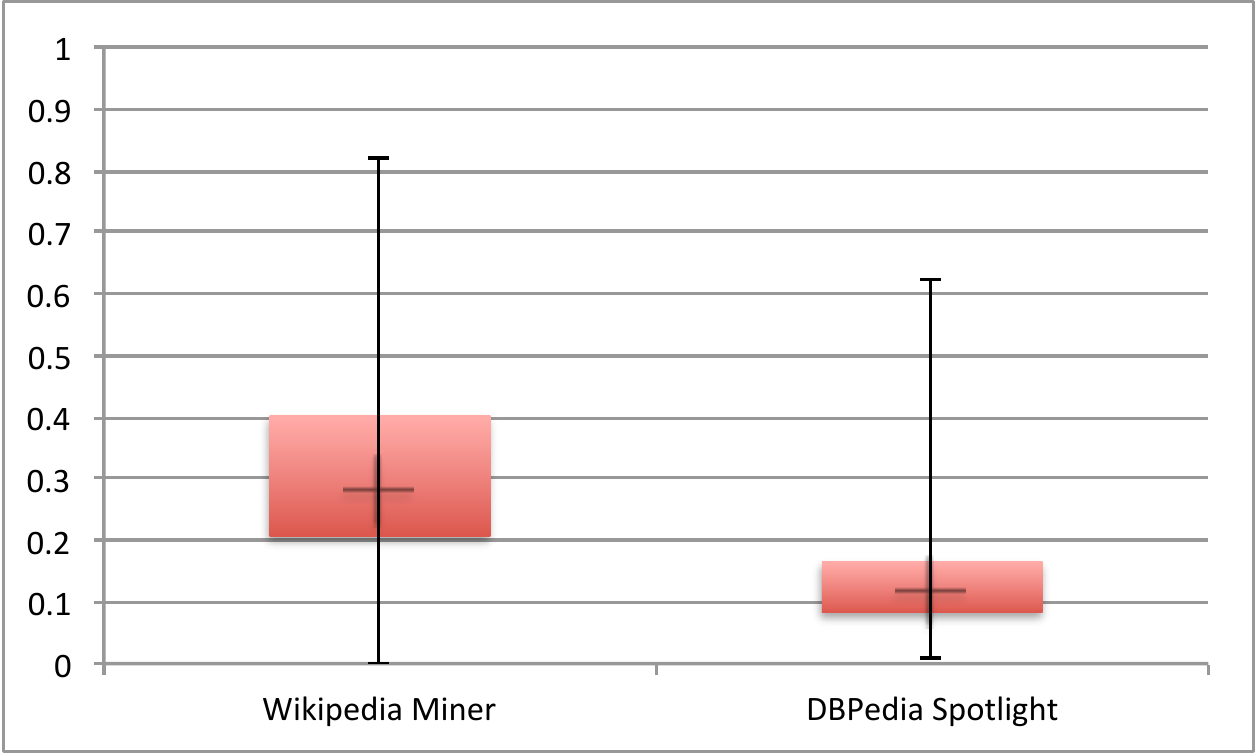}%
}
	\caption{Confidence scores for top candidate}
	\label{fig:confidence}
\end{figure}

A final observation is that not only do ambiguous entities appear frequently,
but the degree of their ambiguity is high as well. Wikipedia Miner and DBPedia
Spotlight assign each candidate meaning a probability score indicative of the
likelihood it is the correct sense. Figure~\ref{fig:confidence}
shows the distribution of this score for each top ranked candidate.
We see the confidence the services are able to assign their selected
candidates is very low, signaling that unreliable knowledge underlies
the ranking. The confidence scores of many correct candidates actually fall
below the default threshold and would erroneously not even be retrieved.
Also, not only do established techniques
have difficulty selecting a candidate with high probability, but there are often
many candidates from which to choose (in our dataset, this number ranges from 2
to 163 with an average of 5-6 and median of 4). All together, this makes the NER
services' candidate selection akin to a random guess with low chance of
choosing the actual intended meaning.

Overall, this analysis confirms entity ambiguity in
short texts is a profuse, difficult problem requiring
attention.

\section{Results}
\label{sec:results}
In this section, we detail the evaluation and performance analysis
we conducted in order to compare the correctness of our ranking algorithms to a
gold standard of human annotator judgments, a number of baseline measures, and
existing state-of-the-art techniques. We evaluated our user-centric approach
with short texts drawn from three separate sources (Twitter, YouTube, and
Flickr) since inspection of posts on these sites revealed qualitative
differences. Table~\ref{tab:performance} reports performance as precision at
rank 1 (P@1), or the fraction of evaluated entities for which the top ranked candidate is the correct meaning; here this
measure is the same as recall since we assume the NER services provide us
every potential candidate sense.
\subsection{Human Annotated Ceiling}
As explained in Section~\ref{sec:data_proc}, in order to obtain labeled data for
an ambiguous entity, we used Mechanical Turk to determine whether or not a
candidate was the correct meaning of an ambiguous entity. We see an average observed agreement across all
coders and items of 0.866 and average Fleiss Kappa=0.803. Both are within
generally acceptable ranges for this type of task, indicating it is
possible to disambiguate these texts manually and feasible to try it
mechanically.
\subsection{Comparison to Baselines}
\begin{itemize}[itemsep=3pt,topsep=3pt,parsep=1pt,partopsep=0pt,leftmargin=2em]
\item Random Candidate (RC) ranks in random order \item Prior Frequency (PF)
ranks candidates according to their commonness, i.e., their prior frequency
measure \item RESLVE with Random-User (RU) applies our method with a random
Wikipedia user's interest model as input rather than that of the user who
uttered the entity \item Wikipedia Miner (WM) and DBPedia Spotlight (DS) are
established NED techniques
\end{itemize}
\begin{table}[b]
	\caption{Precision (P@1) of ranking methods}
	\vspace*{-.4cm}
  	\centering
  	\resizebox{0.75\columnwidth}{!}{
  		\begin{tabular}{lccc}\toprule
      		 & Flickr & Twitter & YouTube \\ \otoprule%
      RESLVE & 0.63 & \textbf{0.76} & \textbf{0.84} \\ \midrule%
      RC & 0.21 & 0.32 & 0.31 \\ \midrule%
      PF & 0.74 & 0.69 & 0.66 \\ \midrule%
      RU & 0.51 & 0.71 & 0.78 \\ \midrule%
      WM & \textbf{0.78} & 0.58 & 0.80 \\ \midrule%
      DS & 0.53 & 0.67 & 0.63 \\ \bottomrule%
  		\end{tabular}
  	}
  	\label{tab:performance}
\end{table}
\subsection{Impact of Text Nature and Length}
Evaluation shows that incorporating background knowledge
about user-specific interests can outperform traditional strategies that rely on
prior word frequencies or linguistic relationships among words in the local
context.
In particular, RESLVE performs best on YouTube, the longest texts in our dataset
(see Figure~\ref{fig:corpora}), mainly because of content-based similarity with
the user-interest model. It also outperforms existing NER services on Twitter
texts, which are generally more personal than those on Flickr or YouTube and
therefore less often simply refer to the most common sense. This shows that
considering user interest for NED can be effective in highly user-centric domains.

Along the
same lines, RESLVE with a random user's interest model as input can perform
better than baselines since incorporating external data
allows more topic overlap with candidate entities; but it is not as
accurate as with the personalized input, showing that user-specific data
does help.

Conversely, RESLVE is less effective on more impersonal text. Misrankings result
from automated posts; and we see lower performance on Flickr data, where many
entities refer to non-subjective topics (e.g., geographic places), which have
high prior frequencies and can be ably resolved with traditional approaches.
Also, examination reveals many of these mentioned places are non-local,
non-familiar travel locations and therefore not covered by a
user's knowledge base contributions about acquainted topics.
Table \ref{tab:errors} and the next section explain additional error cases.
\subsection{Error Analysis and Future Work}
\subsubsection{Incorrect Candidate Selection}
Analyzing RESLVE's failure points reveals reasons for ranking mistakes
and ideas for future improvement.
\begin{table}
	\caption{Errors made by RESLVE}
	\vspace*{-.4cm}
  	\centering
  	\begin{tabular}{p{1.6cm} p{5cm}}\toprule
      	\centering Example & \centering Reason \tabularnewline
      	\toprule
      	``\textbf{\textit{Peter}} on the dock'' & Referent entity not in
      	Wikipedia, so overlap with user model impossible \\ \midrule%
      	``I uploaded a \textbf{\textit{video}} on @youtube'' & RESLVE
      	assumes every entity is user-specific and always selects candidate with
      	most user model overlap, here ranking ``1945 European Films" highest for
      	``video" for a user who edits European cinema pages \\ \midrule%
      	RESLVE assigns correct candidate a 0.0 score  & Performance depends
      	on amount of content user produces in the knowledge base and drops with
      	users who edit shorter articles or make fewer contributions to Wikipedia
      	overall \\ \bottomrule%
  	\end{tabular}
  	\label{tab:errors}
\end{table}
\subsubsection{User Identity}
\label{sec:bridging_results}
The identity module (see Figure~\ref{fig:system_framework}) simply performs
string matching to determine whether a username exists on both a social
media site and on Wikipedia, and Table~\ref{tab:username_reuse} summarizes the
results from our dataset.
\begin{table}[b]
	\caption{Usernames reused}
	\vspace*{-.4cm}
	\centering
  	\begin{tabular}{lcc}\toprule
      		  & \# Usernames & Exist on Wikipedia \\ \otoprule%
      Twitter & 479 & 46.1\% \\ \midrule%
      YouTube & 454 & 19.6\% \\ \midrule%
      Flickr  & 226 & 21.7\% \\ \bottomrule%
  	\end{tabular}
  	\label{tab:username_reuse}
\end{table}
Comparing public profile information between the social media accounts
and Wikipedia accounts that do exist and share the same username, annotators
unanimously found that Twitter-Wikipedia, YouTube-Wikipedia, and
Flickr-Wikipedia username matches actually belong to the same person in
approximately 47\%, 48\%, and 71\% of cases, respectively. In the remaining cases, annotators
were generally unable to make any determination about user identity rather than
confirm a username belonged to different people. While our simple approach
succeeds in more cases than it fails, it does face limitations, and the relatively high costs of determining confirmed
cross-platform user identities limits the sample size of our current evaluation.
Improvement of this module is our first priority for future research. We
identify 3 main sources of errors:
\begin{enumerate}[itemsep=3pt,topsep=3pt,parsep=1pt,partopsep=0pt,leftmargin=2em]
  \item \textit{False positives}: String matching succeeds in retrieving
  accounts that exist on both the social Web and Wikipedia with the same
  username, but these accounts do not belong to the same individual person.
  \item \textit{True negatives}: The individual who participates on the social
  Web does not hold an account on Wikipedia.
  \item \textit{False negatives}: Accounts exist, but a string matching approach
  fails because they have different usernames.
\end{enumerate}

To address the last case where a user contributed to a knowledge base but
under a different username, we propose exploring a model that considers a larger
array of fully and partially matching profile information. Recent work finds profile attributes such as name, email address, or
hometown to be highly reliable identifiers \cite{abel2013user,
Carmagnola:2009ij, Vosecky:2009kp}.

To address the opposite case where the user has not contributed to a knowledge
base, we propose collaborative filtering techniques to use
the contribution histories of social connections (people a user
friends, follows, or favorites on the social Web) as an approximation of the
user's own interests. Extending the implementation to
other knowledge bases besides Wikipedia is also possible, as is
modeling user interest not only from contributions like article edits but also
from other forms of participation such as page visits, favoriting, or
bookmarking.
\section{Conclusion}
\label{sec:conclusion}
We addressed the Named Entity Disambiguation problem with a user-interest
centered approach. We showed that a user tends to produce content within a scope
of topics of personal interest across multiple online platforms and
that it is possible to formally represent these topics using structured semantic
data that can serve as a personal knowledge context. We introduced a
representation of user interest given any general knowledge base and implemented
the model on Wikipedia as a popular and powerful instance.

Our approach to the NED problem does not depend on local or
language-specific information, which is often hard to process, unreliable, or
missing entirely in user-generated content. We reveal the advantages of our
strategy over a variety of baseline and state-of-the-art methods and achieve
improvements in performance, especially
when text contains content of a highly personal nature.

As the Web and particularly the social Web evolves, this kind of
content will only continue to extensively grow, and user profiles will become
ever more inter-connected across domains. Not only will this compel innovative
solutions to handle existing information retrieval tasks, but as these tasks
become increasingly difficult, it will make the solutions introduced in this
paper all the more feasible and effective.
\balance
\section{Acknowledgements}
This material is based upon work supported by the National Science
Foundation Graduate Research Fellowship under Grant No. DGE 1144153, and this
work is also supported by a Marie Curie International Outgoing Fellowship within
the 7th European Community Framework Progra\-mme (PIOF-GA-2009-252206).


\bibliographystyle{abbrv}
\bibliography{references}
%
%
\end{document}